\documentclass[aip,amssymb,amsmath]{revtex4-1}

\usepackage{docs}%
\usepackage{graphicx}
\usepackage{dcolumn}
\usepackage{bm}
\renewcommand

\usepackage{here}

\usepackage{hyperref}
\hypersetup{pdfstartview={FitH},pdfpagemode={UseNone},
            colorlinks,linkcolor=blue, citecolor=blue, urlcolor=blue,
            bookmarksopen=true, pdfnewwindow=true}
\usepackage[all]{hypcap}

\graphicspath{{pict/}{subm/}{}}

\newcommand\reqtn[1]{\ref{eq:#1}}
\newcommand\reqt[1]{(\reqtn{#1})}

\newcommand\lsect[1]{\protect\label{sect:#1}}
\newcommand\rsect[1]{\ref{sect:#1}}

\newcommand\hrefBibPDF[3][]{}

\newcommand{\ket}{\rangle}

\expandafter\ifx\csname package@font\endcsname\relax\else
 \expandafter\expandafter
 \expandafter\usepackage
 \expandafter\expandafter
 \expandafter{\csname package@font\endcsname}%
\fi

\usepackage{braket}
\begin{document}
\sloppy
\title{Synthetic photonic lattice for single-shot reconstruction of frequency combs}

\author{James G. Titchener}
\affiliation{Nonlinear Physics Centre, Research School of Physics, The Australian National University, Canberra, ACT 2601, Australia}
\affiliation{Quantum Technology Enterprise Centre, Quantum Engineering Technology Labs, H. H. Wills Physics Laboratory and Department of Electrical and Electronic Engineering, University of Bristol, BS8 1FD, UK}

\author{Bryn Bell}
\affiliation{Institute of Photonics and Optical Science (IPOS), School of Physics, University of Sydney, Sydney, NSW 2006, Australia}
\affiliation{QOLS, Department of Physics, Imperial College London, London SW7 2AZ, UK}

\author{Kai Wang}%
\thanks{Present address: Ginzton Laboratory, Stanford University, Stanford, CA 94305, USA}
\affiliation{Nonlinear Physics Centre, Research School of Physics, The Australian National University, Canberra, ACT 2601, Australia}

\author{Alexander~S.~Solntsev}
\affiliation{School of Mathematical and Physical Sciences, University of Technology Sydney, 15 Broadway, Ultimo NSW 2007, Australia}
\affiliation{Nonlinear Physics Centre, Research School of Physics, The Australian National University, Canberra, ACT 2601, Australia}

\author{Benjamin~J.~Eggleton}%
\affiliation{Institute of Photonics and Optical Science (IPOS), School of Physics, University of Sydney, Sydney, NSW 2006, Australia}

\author{Andrey A. Sukhorukov}%
\email{andrey.sukhorukov@anu.edu.au}
\affiliation{Nonlinear Physics Centre, Research School of Physics, The Australian National University, Canberra, ACT 2601, Australia}

\date{\today}

\begin{abstract}
We formulate theoretically and demonstrate experimentally an all-optical method for reconstruction of the amplitude, phase and coherence of frequency combs from a single-shot measurement of the spectral intensity. Our approach exploits synthetic frequency lattices with pump-induced spectral short- and long-range couplings between different signal components across a broad bandwidth of of hundreds GHz in a single nonlinear fiber. When combined with ultra-fast signal conversion techniques, this approach has the potential to provide real-time measurement of pulse-to-pulse variations in the spectral phase and coherence properties of exotic light sources.
\end{abstract}
\maketitle

\section{Introduction}

The concept of synthetic frequency lattice in photonics~\cite{Yuan:2018-1396:OPT, Ozawa:2019-349:NRP}, where dynamic modulation induces cross-talks between discrete spectral lines to produce lattice-like behavior~\cite{Bell:2017-1433:OPT, Dutt:2020-59:SCI}, is stimulating rapidly growing research in recent years. The development of synthetic lattices not only provides a new means for the reshaping of discrete spectral lines, but also enables tailored transformations of optical spectra underpinned by the remarkable capability of inducing multiple arbitrary short- and long-range couplings in frequency domain~\cite{Bell:2017-1433:OPT}, beyond the physical constraints of spatial lattices~\cite{Szameit:2009-2838:OL}. So far, synthetic lattices were extensively investigated in their capacity for creating multiple artificial dimensions~\cite{Yuan:2016-741:OL, Ozawa:2016-43827:PRA, Lustig:2019-356:NAT, Maczewsky:2020-76:NPHOT, Wang:2002.08591:ARXIV}, yet their other potential applications remain largely undeveloped. This motivates our current research on employing the  fundamental advances of synthetic lattices to the practically important task of fast and efficient characterisation of frequency combs.

The full characterization of ultra-short optical signals, including both their phase and coherence properties, is crucial for the development and understanding of novel engineered light sources, such as optical frequency combs~\cite{Walmsley:2009-308:ADOP, Kippenberg:2011-555:SCI, Schmeissner:2014-263906:PRL}, frequency encoded quantum states \cite{Kues:2017-622:NAT}, and optical soliton molecules~\cite{Herink:2017-50:SCI}. Furthermore, full optical signal characterization is important for the communication of timing information over fiber networks \cite{Foreman:2007-21101:RSI} and wavelength division multiplexing transmission formats where the relative phases between individual carriers are important~\cite{Ellis:2005-504:IPTL}.
The most commonly used methods for measuring optical pulses, frequency-resolved optical gating (FROG) \cite{Scott:2007-9977:OE} and Spectral Phase Interferometry for Direct Electric-field Reconstruction (SPIDER) \cite{Walmsley:2009-308:ADOP,Dorrer:2003-477:OL}, require complex multimode optical setups in order to reconstruct the amplitude and degree of coherence.

Techniques with the capability to recover spectral phase information with just a single spatial mode are being actively developed. This includes
ultra-fast signal conversion methods like stretch transform spectroscopy~\cite{Solli:2008-48:NPHOT}, which could allow real time measurement of pulse to pulse variations in the spectral phase profile, without a need to use an interferometric setup~\cite{Xu:2016-27937:SRP}. The technique of electro-optic spectral shear interferometry (EOSI)~\cite{Wong:1994-287:OL, Dorrer:2003-477:OL, Bromage:2006-3523:OL} can also be implemented for determining the spectral phase and amplitude of frequency combs using a single mode setup, by employing an electro-optic modulator (EOM)~\cite{Supradeepa:2010-18171:OE}. The electro-optic phase modulation enables particularly effective characterization of weak coherent frequency combs, 
even in quantum regime at the single- and few-photon levels~\cite{Kues:2017-622:NAT, Imany:2018-13813:PRA}. However, EOSI necessarily requires a sequence of different EOM shears to be applied to the signal to fully determine its state, which significantly limits the data acquisition speed.
Furthermore, the temporal resolution of EOSI is limited by the free spectral range of the EOM, while
the steady growth of telecommunication bandwidth calls for fast and efficient methods of optical spectral characterisation.

Here we propose and experimentally demonstrate that specially designed synthetic lattices can facilitate a new method for recovering the full spectral amplitude, phase and coherence information of an optical frequency comb from a single-shot measurement of the spectral intensities after an input signal is transformed in a nonlinear fiber. This is mediated by a co-propagating pump, which is tailored to realize an effective 
synthetic lattice for the signal through the energy-conserving process of four-wave mixing Bragg scattering (FWM-BS)~\cite{Bell:2017-1433:OPT}.
The presented method can be much faster than FROG, SPIDER or EOM based approaches, since it only needs a single measurement and minimal post-processing, while allowing for a significantly larger free spectral range compared to the EOM approach.

In the following, we first introduce the concept and theory of the reconstruction with a synthetic frequency lattice in Sec.~\rsect{concept}. Then, we describe the experimental setup and demonstrate the reconstruction of signals with two and four spectral lines in Sec.~\rsect{experiment}. Finally, we present conclusions and outlook in Sec.~\rsect{conclusions}.

\section{Concept of reconstruction with synthetic frequency lattice}
\lsect{concept}
We propose to use a synthetic frequency lattice to transform input signals into a form where a single-shot spectral intensity measurement at the output would allow full reconstruction of the input spectral phase and coherence. In a synthetic lattice, discrete frequency components effectively become lattice sites, where light can couple or hop to nearest-neighbor or farther frequencies corresponding to the short- and long-range couplings on a lattice.
Such frequency lattices allow complex spectral manipulation of light within a single spatial mode of an optical fiber or waveguide~\cite{Yuan:2016-741:OL, Bell:2017-1433:OPT, Bersch:2009-2372:OL}.

We build on the recently developed approaches based on spatial lattices~\cite{Minardi:2012-3030:OL, Minardi:2015-13804:PRA, Titchener:2016-4079:OL, Oren:2017-993:OPT, Titchener:2018-19:NPJQI,  Wang:2019-41:OPT}, and reveal that light propagation and reshaping 
in nonlinearly induced synthetic lattices can be utilized to allow full reconstruction of an input signal from a simple spectral intensity measurement at the output. Specifically, the spectral amplitude, phase and relative coherence of the signal can be recovered. 
The system is all-optical, with 
the synthetic lattice induced through optical nonlinearity and controlled via a pump, which  
can be altered quickly to provide different functionalities.  Furthermore, manipulation of light within the spectral dimensions, without resorting to spatial multiplexing, makes such devices very compact.
\begin{figure*}
	\centering
	\includegraphics[width=\textwidth]{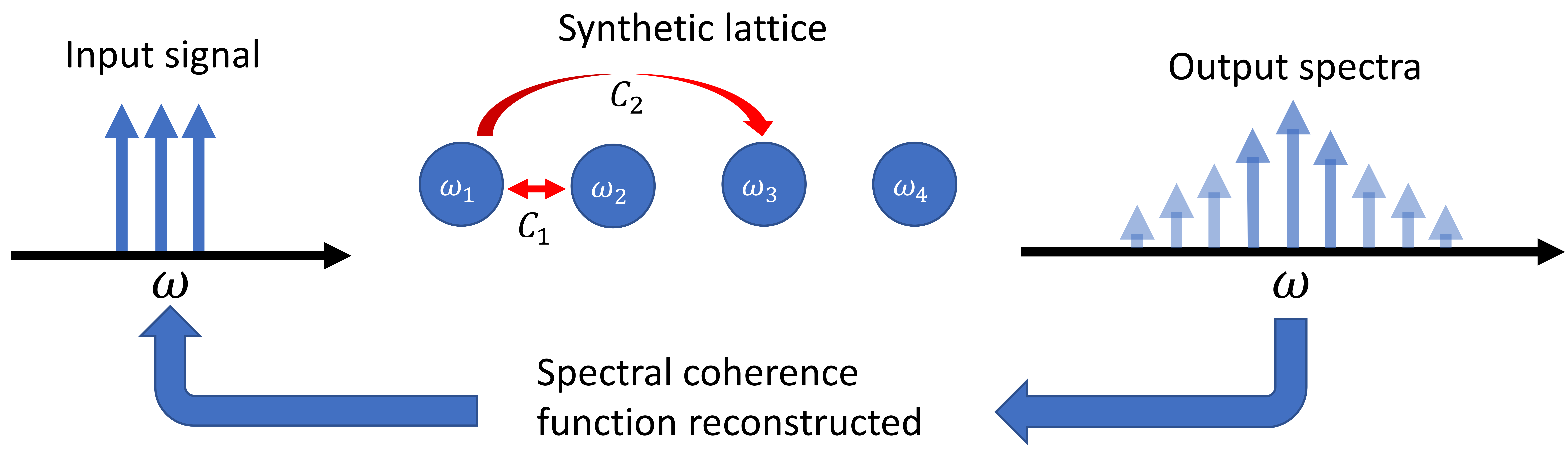}
	\caption{Concept of signal reconstruction.
		An input signal composed of a number of different frequency channels undergoes a transformation in a synthetic lattice with first and second order coupling between frequency channels. The observed output spectra can then uniquely determine the input spectral coherence function of the signal for a specially tailored form of the synthetic lattice transformation.   }
	\label{fig:1}
\end{figure*}

We consider an input signal which spectrum consists of a number of discrete frequency components separated by a constant spectral spacing $\Omega$, as shown in Fig.~\ref{fig:1}. 
After propagating through the synthetic lattice, which introduces coupling between the frequency lines, the input signal is spread across an increased number of synthetic lattice sites.
In the regime of coherent frequency conversion, the output spectral intensity pattern is dependent on the relative phases and coherence of the input signal. Importantly, we find that, for certain parameters of the synthetic lattice,
it is sufficient to measure just the output spectral intensities in order to perform robust reconstruction of the input spectrum and phase.
The propagation of a fully coherent signal through the synthetic lattice, which can be induced experimentally in an optical fiber as we discuss in the following Sec.~\rsect{experiment}, can be described by coupled-mode equations~\cite{Bell:2017-1433:OPT} for a set of complex amplitudes $a_n$ at frequency components $n \Omega$, 
\begin{equation}\label{eq:hamiltonian}
   \frac{\mathrm{d}a_n}{\mathrm{d} z}=i \sum_{j=1}^{+\infty} \left[ C_j a_{n+j}+C^\ast_j a_{n-j}  \right],
\end{equation}
where $C_j = 2\gamma\sum_{m} A_m(0) A_{m-j}^\ast(0)$ is the $j$-th order coupling coefficient, $\gamma$ is the effective nonlinearity, and $A_m(z)$ are the complex amplitudes of the envelope of pump spectral components.
We determine the lattice dispersion in the framework of  Eq.~(\ref{eq:hamiltonian}) by considering the plane-wave solutions in the form $a_n = a_0 \exp[i k n + i \beta(k) z]$, where $k$ is the wavenumber, and find the propagation constant as:
\begin{equation}\label{eq:beta}
    \beta(k) = 2 \sum_{j=1}^{+\infty} {\rm Re}(C_j e^{i k j}) .
\end{equation}

The output state after propagation through the fiber length $L$ is 
related to the input state via a linear transformation which can be expressed through the transfer matrix ${\bf T}$, 
\begin{equation}\label{eq:output}
   a_n(z=L) = \sum_i {\bf T}_{n,n'}(L) a_{n'}(z=0).
\end{equation}
We obtain the transfer matrix elements by performing a Fourier transform of Eq.~\reqt{hamiltonian} as
\begin{equation}\label{eq:Tn}
   {\bf T}_{n,n'}(z) = \frac{1}{2 \pi} \int_{-\pi}^{\pi} dk \exp\left[ i k (n-n') + i \beta(k) z \right] .
\end{equation}

In general, noise and fluctuations may affect the coherence of comb spectrum. Such partially coherent signals can be characterized by the mutual coherence or complex visibility~\cite{Mandel:1995:OpticalCoherence} between different pairs of spectral lines,
\begin{equation}\label{eq:visibility}
   V_{n,m}(z) = \langle \tilde{a}_n(z) \tilde{a}_m^\ast(z) \rangle ,
\end{equation}
where $\tilde{a}_n(z)$ are the fluctuating complex field amplitudes of the spectral comb lines and angled brackets denote averaging. It is established~\cite{Mandel:1995:OpticalCoherence} that any partially coherent field can be represented as an incoherent mixture of coherent fields $a_m^{(p)}$ indexed by $p$, such that
\begin{equation}\label{eq:visibilityMixture}
   V_{n,m}(z)  = \sum_p a_n^{(p)}(z) [a_m^{(p)}(z)]^\ast .
\end{equation}
We substitute Eq.~\reqt{output} into Eq.~\reqt{visibilityMixture} and obtain
\begin{equation}\label{eq:visibilityTransfer}
   V_{n,m}(z)  = \sum_p \sum_{n',m'} {\bf T}_{n,n'}(z) a_{n'}^{(p)}(0) \left[{\bf T}_{m,m'}(z) a_{m'}^{(p)}(0)\right]^\ast = \sum_{n',m'} {\bf T}_{n,n'}(z) {\bf T}_{m,m'}^\ast(z) V_{n',m'}(0).
\end{equation}
The output comb intensities, which can be directly measured by a spectrometer, are then found as
\begin{equation}\label{eq:forwards_eq}
   I_{n}(z=L) 
   \equiv V_{n,n}(L) 
    =\sum_{n',m'}
    T_{n,n'}(L)  T_{n,m'}^\ast(L) \, V_{n',m'}(z=0) .
\end{equation}
Therefore the input complex visibility can be retrieved
by inverting Eq.~(\ref{eq:forwards_eq}) to find $V_{n',m'}(z=0)$ in terms of the output spectral intensities $I_{n}(z=L)$. Considering the input signal to consist of $N_{\rm in}$ spectral lines, and the output spread out to $N_{\rm out}$ lines, we require that the number of measurements is equal to or larger than the number of unknowns, $N_{\rm out} \ge N_{\rm in}^2$.
The robustness of the inversion process can be characterized by the ``condition number'', $\kappa=||\mathbf{M}||\; ||\mathbf{M}^{-1}||$, where $\mathbf{M}$ is the instrument matrix that directly links the observable ($I_n(z=L)$) and the reconstructed quantities ($V_{n',m'}(z=0)$) and $||\cdot||$ denotes the norm of a matrix.
The condition number quantifies the amplification of the measurement errors and noise of the output intensities in the process of reconstruction of the input
complex visibility function~\cite{Press:2007:NumericalRecipes, Minardi:2015-13804:PRA, Titchener:2018-19:NPJQI, Wang:2018-1104:SCI}. A high condition number implies that small errors in the measurement data would be highly amplified after the inversion, making the result of reconstruction unreliable. Thus for an accurate reconstruction, we require a low condition number.
As we discuss in the following, robust reconstruction, associated with a small condition number close to a theoretical optimum, requires specially tailored simultaneous short- and long-range couplings in the synthetic frequency lattice. We note that the reconstruction procedure is also possible in the presence of absorption, which effect can be incorporated in the coupled-mode Eq.~(\ref{eq:hamiltonian}), while the general approach formulated in Eqs.~(\ref{eq:output}),(\ref{eq:visibility})-(\ref{eq:forwards_eq}) remains the same.

\section{Experimental comb reconstruction via the synthetic lattice in an optical fiber} \lsect{experiment}

\begin{figure*}
	\centering
	\includegraphics[width=.8\textwidth]{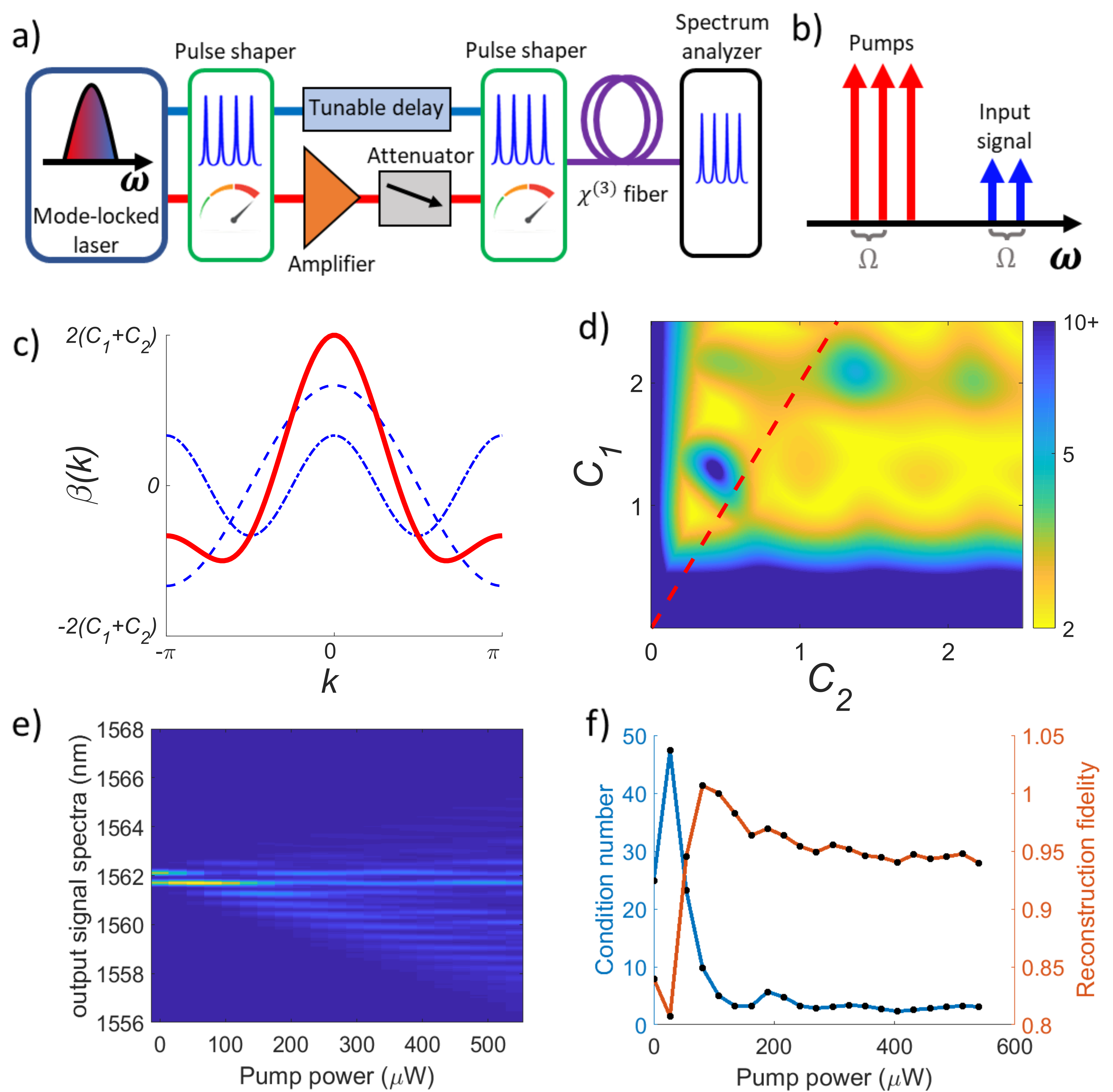}
	\caption{
		(a)~Experimental setup. A mode locked fiber laser is spectrally shaped to create a set of pump (red) and signal (blue) spectral lines.
        (b)~Example pump and signal spectra after being shaped by the pulse shapers and before coupling into the nonlinear fiber in (a).
		(c)~The dispersion of the synthetic lattice shown with the red line. The dashed lines show the contributions of the first- and second-order coupling to the dispersion, for $C_1 = 2 C_2$.
		(d)~The condition number of the transformation as a function of first and second order coupling rates. The red dashed line shows the subset of the parameter space with $C_1 = 2 C_2$ which was investigated experimentally.
		(e)~Experimentally measured output spectra for a frequency space quantum walk for input state $\ket{0}+i\ket{1}$.
		(f)~(left hand axis) Experimental condition number of the quantum walk transformation shown in~(e). Increasing the pump power increases the coupling rates, effectively moving along the red line in Fig.~\ref{fig:2}(d).
		(right hand axis) Fidelity of the reconstruction of the state $\ket{0}+i\ket{1}$ as a function of pump power.}
	\label{fig:2}
\end{figure*}

We realize the synthetic frequency lattice through the process of four-wave mixing Bragg scattering (FWM-BS) in a $\chi^{(3)}$ nonlinear fiber \cite{Bell:2017-1433:OPT}, see Fig.~\ref{fig:2}(a). Here, the lattice sites are represented by discrete signal frequency channels, which are coupled together by nonlinear frequency conversion. In this system, controllable coupling is made possible by shaping the spectrum of the optical pump. A broad phase-matched bandwidth for FWM-BS is achieved  by placing the signal and pump spectra on either side of the zero-dispersion wavelength. One of the unique features of this approach is that it allows long-range coupling, which enables efficient reconstruction of complex input signal states by measuring the output spectra.

We experimentally demonstrate the reconstruction of a signal consisting of up to four frequency channels, but in principle the method is applicable to more complicated signals. 
We choose a pump spectrum consisting of $3$ different frequency components around $1540$~nm wavelength, separated by $\Omega = 2\pi \times 50$\,GHz [Fig.~\ref{fig:2}(b)]. By having three equally spaced pump spectral lines with $A_1=A_2=A_3$, the signal can be up- or down-shifted by $\Omega$ or $2\Omega$ via the FWM-BS. This realizes a synthetic lattice with first and second order coupling coefficients $C_1 = 2 C_2 = 4 \gamma |A_1|^2$. The corresponding lattice dispersion $\beta(k)$ is presented in Fig.~\ref{fig:2}(c), with the contributions of the first and second order coupling shown individually as dashed lines.

We first numerically 
investigate the inversion problem for the case where the input signal is limited to two input synthetic lattice sites. We study the effect of varying the coupling rates $C_1$ and $C_2$, while assuming that other couplings are zero ($C_{j>2}\equiv 0$),
and the normalized fiber length is held constant at $L=1$. For each set of values of $C_1$ and $C_2$ we calculated the transfer function according to Eq.~(\ref{eq:Tn}) and the corresponding condition number. 
The plot of the condition number vs. the coupling rates is shown in Fig.~\ref{fig:2}(d), indicating that inversion will be practical given $C_1$ and $C_2$ are chosen in an optimal region. Importantly, both short- and long-range couplings need to be present simultaneously, with $C_{1,2} > 0$. 

Next, we experimentally measure the transfer function of the nonlinear fiber system for a range of different coupling rates. This measurement fully incorporates the effect of losses both on the pump and signal, which enables reconstruction under realistic experimental conditions.
We vary the 
power of the three equal-amplitude pump spectral lines $A_1=A_2=A_3$ to simultaneously control the first- and second-order coupling rates with $C_1=2 C_2$, as indicated by the dashed line in Fig.~\ref{fig:2}(d).
The transfer function of the system for a given set of pump powers was determined by performing measurements for a set of four different input signals. We used the basis set $\ket{\omega_0}$, $\ket{\omega_1}$,  $\ket{\omega_0}+\ket{\omega_1}$ and $\ket{\omega_0}+i\ket{\omega_1}$, where $\omega_0$ and $\omega_1$ correspond to the wavelengths of $1561.7$~nm and $1562.1$~nm, respectively. The output states resulting from the input basis state $\ket{\omega_0}+i\ket{\omega_1}$ are shown in Fig.~\ref{fig:2}(e), where the x-axis indicates the pump power.
We use the experimentally determined transfer function to calculate the condition number of the inverse problem as a function of pump power, see Fig.~\ref{fig:2}(f). We also solved the inverse problem using the data from the basis state $\ket{\omega_0}+i\ket{\omega_1}$ as the input, then calculated the fidelity of the reconstructed state with the input. This fidelity is shown by the right hand y-axis on Fig.~\ref{fig:2}(f). The decrease in fidelity with increasing pump power indicates that the assumption that the fiber system preforms  linear transformation on the input signal breaks down, as other effects begin to emerge at higher pump powers. We use Fig.~\ref{fig:2}(f) to determine the pump power which minimizes the condition number and maximizes the fidelity in order to allow the most accurate reconstruction of arbitrary input states.

\begin{figure*}
	\centering
	\includegraphics[width=\textwidth]{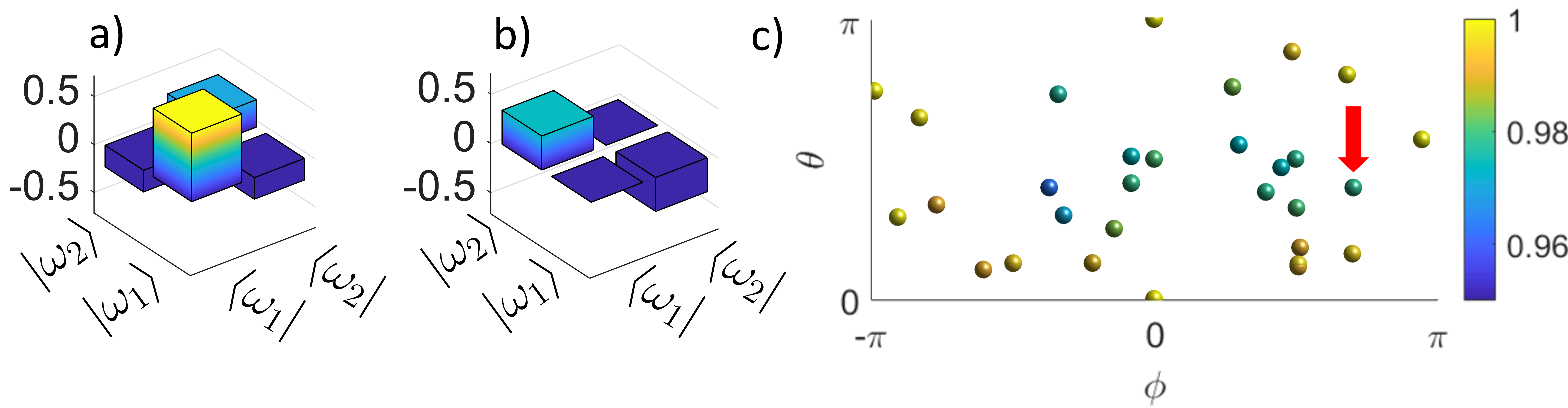}
	\caption{
		(a)~Real and (b)~imaginary parts of an example reconstructed two-channel spectral coherence function. Fidelity to the target state is $97.8\%$. (c)~Reconstruction fidelity shown by color of random states in the form $\cos(\theta/2)\ket{\omega_0}+e^{i\phi}\sin(\theta/2)\ket{\omega_1}$ vs. the parameters $\phi$ and $\theta$ as indicated by labels.
		The red arrow points to a state from (a) and (b).}
	\label{fig:3}
\end{figure*}

A pump power of $135\:\mu W$ was chosen to enable the optimal reconstruction accuracy of input signals. We then tested our  reconstruction method experimentally by coupling a sequence of random input signals into the fiber. The general form of the input signals was $a_n = \cos(\theta/2)\ket{\omega_0}+e^{i\phi}\sin(\theta/2)\ket{\omega_1}$, where $\theta$ and $\phi$ are randomly chosen, and implemented using a waveshaper. Then the output spectra at the end of the fiber are measured and used to reconstruct the input signal. An example of the real and imaginary parts of a reconstructed input complex visibility function is shown in Figs.~\ref{fig:3}(a) and~\ref{fig:3}(b), respectively. The reconstructions were carried out for many random input signals parametrized by $\theta$ and $\phi$ as shown in Fig.~\ref{fig:3}(c). We observe that the reconstructed signals' fidelity to the input that the waveshaper was configured to produce is typically above $96\%$.

\begin{figure*}
	\centering
	\includegraphics[width=\textwidth]{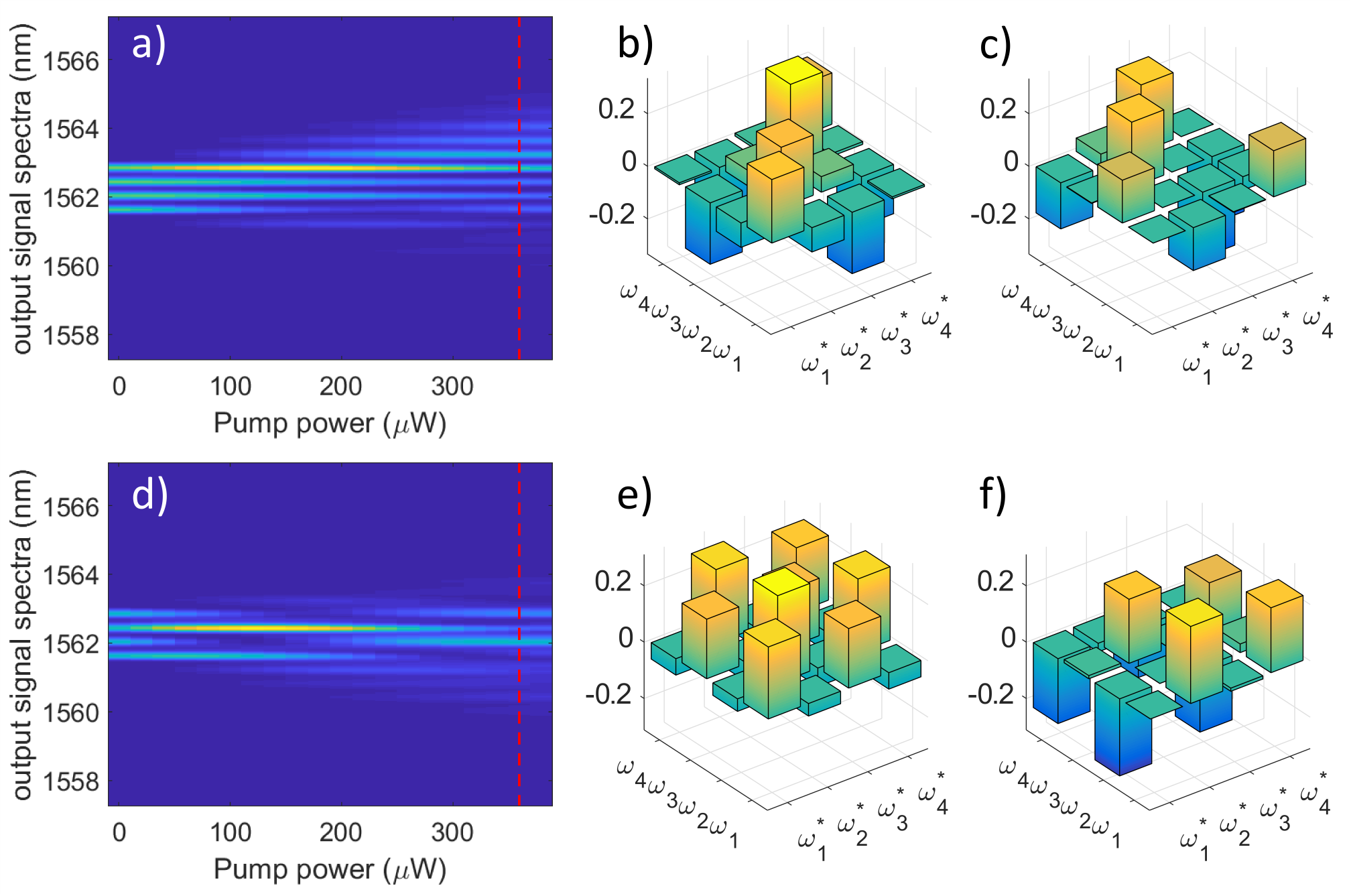}
	\caption{
		(a,d)~Output spectra after a quantum walk in the synthetic lattice as a function of the pump power for the input state (a)~$\ket{\omega_1}+i\ket{\omega_2}-\ket{\omega_3}-i\ket{\omega_4}$ and (d)~$i\ket{\omega_1}+\ket{\omega_2}+i\ket{\omega_3}+\ket{\omega_4}$.
		(b,e)~Real and (c,f)~imaginary parts of the spectral coherence function, reconstructed from the the output spectra measured at $360\:\mu W$ pump power [red dotted line in (a,d)] as shown on the corresponding plots on the left.
		Fidelity of the reconstruction in (b,c)~is $93.9\%$ and in (e,f)~$96.7\%$.
	}
	\label{fig:4}
\end{figure*}

We also demonstrate the reconstruction of more complex input signals. As an example, we consider an input consisting of four lattice sites centered at frequencies $1561.6$~nm, $1562.0$~nm, $1562.4$~nm, and $1562.8$~nm, which we denote by the kets $\ket{\omega_1}$, $\ket{\omega_2}$, $\ket{\omega_3}$, and $\ket{\omega_4}$, respectively. First the transfer function of the fiber and pump system was determined by measuring the output spectra of a set of 16 basis states. Then this was used to reconstruct a number of random input states. The output spectra as a function of pump power of two representative input states are shown in Figs.~\ref{fig:4}(a) and~\ref{fig:4}(d). The pump power providing optimum condition number of 24.7 for inversion was determined to be $360\:\mu W$, as indicated by the dashed vertical lines in Figs.~\ref{fig:4}(a) and~\ref{fig:4}(d). 
The real and imaginary parts of the reconstructed complex visibility functions are shown in Figs.~\ref{fig:4}(b,c) and  \ref{fig:4}(e,f). We obtain a high fidelity of the reconstructed states to the input at $93.9\%$ and $96.7\%$, respectively.

\section{Conclusion and outlook} \lsect{conclusions}

In conclusion, we have experimentally investigated the use of synthetic lattices for single-shot reconstruction of frequency combs.
We have shown that the full complex visibility of input spectral comb signals can be reconstructed just by measuring its output spectra after propagating through a synthetic lattice with specially optimized short- and long-range couplings.
This technique is naturally suited to characterizing input signals consisting of a number of evenly spaced narrow frequency components. Furthermore, there appears an interesting potential to extend the approach for characterizing non-equidistant spectra by using more complex pump profiles to induce couplings with incommensurate frequency spacings.

The developed single-shot nonlinear-optical frequency comb characterisation is also well-suited for weak non-classical light, since the nonlinear process of FWM-BS does not introduce quantum noise. Recently, there has been significant progress in quantum optics in the spectral domain~\cite{Reimer:2016-1176:SCI, Kues:2017-622:NAT}. Frequency domain Hong-Ou-Mandel interference has proven stable and practical to implement~\cite{Kobayashi:2016-441:NPHOT, Imany:2018-2760:OL}, and frequency-encoded photonic states are now poised to enable scalable quantum information processing~\cite{Lukens:2017-8:OPT, Lu:2018-30502:PRL, Lu:2018-1455:OPT}. Having access to fast and simple means to perform a full tomographic characterization of a quantum state in the frequency domain would be indispensable. We anticipate a potential in applying our approach based on synthetic frequency lattices in the quantum regime, where the measurement of output multi-photon correlations between different combinations of the spectral lines can enable full reconstruction of the input quantum density matrix based on the mathematical formalism developed in the spatial domain~\cite{Titchener:2016-4079:OL, Oren:2017-993:OPT, Titchener:2018-19:NPJQI,  Wang:2019-41:OPT}, bringing the benefits of simplicity and speed of all-optical realization.

\section*{ACKNOWLEDGMENTS}

We acknowledge support by the Australian Research Council (ARC): Discovery Project (DP160100619,  DE180100070, and DP190100277); Centre of Excellence CUDOS (CE110001018); Laureate Fellowship (FL120100029).

\end{document}